DETERMINATION AND CONTROL OF OXYGEN STOICHIOMETRY IN THE CUPRATE $Bi_2Sr_2CuO_{6+\Delta}$.

F. JEAN*,§, D. COLSON&, G. COLLIN*, N. BLANCHARD°,* and A. FORGET&

*LLB, DSM / DRECAM, CEA-Saclay 91191 Gif-sur-Yvette Cedex, France
§LEMHE, Université Paris-Sud, Bât.415, 91405 Orsay Cedex, France
&SPEC, DSM / DRECAM, CEA-Saclay 91191 Gif-sur-Yvette Cedex, France
°LPS, Université Paris-Sud, Bât.510, 91405 Orsay Cedex, France

The relatively low Tc of $Bi_2Sr_2CuO_{6+\Delta}$ allows to study normal-state down to low temperatures and the non-substituted compound is intrinsically strongly overdoped. Hole concentration can be adjusted trough oxygen excess, but few data exist in the literature about the quantitative control of $\Delta$. The synthesis, achieved in air by solid-state reaction, needs long-time annealing to obtain pure phase with stoichiometric cationic ratios. Thermogravimetric techniques were used to explore oxygen non-stoichiometry. Absolute oxygen content was determined by reduction with hydrogen, while the oxygen exchange was studied between 300°C and 670°C with $10^{-4} \leq P_{O_2} \leq 1$ atm. Oxygen excess varies between 0.14 and 0.18, with possibly two regimes of oxygen intercalation. These results are compared to Bi-2212.

1. INTRODUCTION

The high-Tc superconducting oxides present oxygen non-stoichiometry, which is closely related to electronic and magnetic properties. Whereas the underdoped region of the phase diagram has been extensively studied, the strongly overdoped regime, currently considered as a conventional Fermi liquid, has been less investigated. However, properties near the suppression of superconductivity are still poorly understood.

The cuprate $Bi_2Sr_2CuO_{6+\Delta}$ (Bi-2201) is an ideal candidate for studying the normal state in this regime. It is intrinsically overdoped, the hole concentration can be adjusted through oxygen excess, and its relatively low critical temperature ($Tc_{max}$~20 K) allows studying the normal state properties over a large temperature range.

Recently, our group has reported anomalous spin susceptibility and thermopower in the strongly overdoped Bi-2201[1,2]. In these experimental studies, relevant conclusions can be drawn if an adapted stoichiometry control is developed[3].

In this work, oxygen non-stoichiometry and thermodynamic properties of $Bi_2Sr_2CuO_{6+\Delta}$ are investigated by thermogravimetric techniques on wide range of oxygen partial pressure (1-10$^{-4}$ atm), in conditions of thermodynamic equilibrium. An attempt is made to clarify the conditions for excess oxygen uptake in a temperature range compatible with the system stability.

2. EXPERIMENTAL

Polycrystalline $Bi_2Sr_2CuO_{6+\Delta}$ was prepared by solid state reaction in air, starting from a stoichiometric mix of oxides ($Bi_2O_3$, $CuO$) and strontium carbonate. Powder diffraction method with CuK$\alpha$ radiation was used to follow the synthesis advancement. Thermogravimetric studies were achieved using three different balances by Setaram (Labsys, B60) and Netzsch (STA449C) with sensibilities from 1 to 1.5 µg, which represent, in composition, much less than 0.001 oxygen per formula unit as our samples were pellets of ~3 g. Reversible oxygen exchange was monitored through temperature between 300°C and 670°C, and atmosphere composition with $10^{-4} \leq P_{O_2} \leq 1$ atm. To determine absolute oxygen content, samples of ~0.5 g were reduced into Bi, Cu and SrO under $H_2$ / Ar flow (10% $H_2$) at 460°C. It should be noted that initial desorption of adsorbed species (~3 mg of $H_2O$ and $CO_2$ per gram of Bi-2201) was observed at 700°C under $P_{O_2}$ = 1 atm for several days at the beginning of every experiment.

3. RESULTS

According to powder x-ray diffraction patterns analysis, pure phase with no significant departure from ideal cationic ratios 2:2:1 was obtained after long-time annealing at 700-730°C[3].

Figure 1 shows typical thermogram when Bi-2201 is reduced by hydrogen, where the measured mass loss yields the oxygen excess in the initial state (300°C, $P_{O_2}$ = 0.1 atm). The experiment was repeated several times and the uncertainty calculated from statistical dispersion, as the confidence interval with a confidence level at 80% : $\Delta = \overline{\Delta} \pm 1.44 \times \sigma/\sqrt{n}$, where $\sigma = [(n \times \Sigma\Delta^2 - (\Sigma\Delta)^2)/(n \times (n-1))]^{1/2}$ and n is the number of experiments. It was found that $\Delta = 0.18 \pm 0.04$ at 300°C in pure $O_2$, and this value, consistent with other determinations[4,5], was taken as a reference for the reversible oxygen exchange measurements.

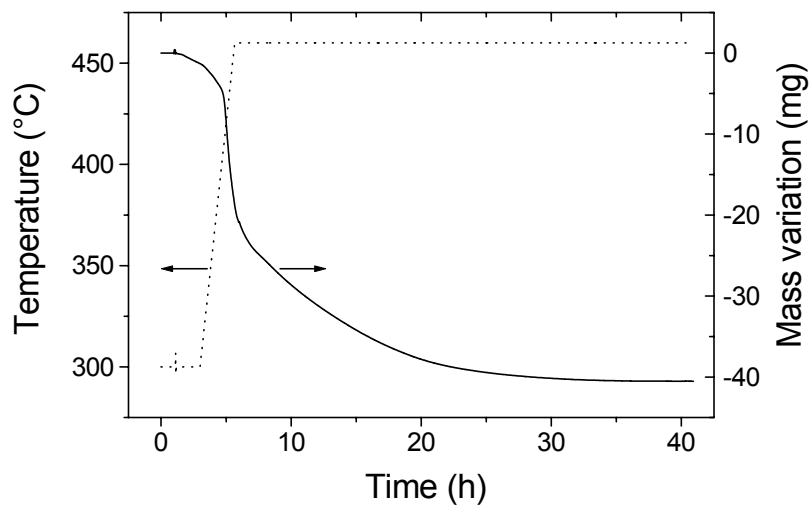

FIGURE 1
Reduction of Bi-2201 by $H_2$. The sample was previously treated at 700°C under $Po_2$ = 0.1 atm to allow $H_2O$ / $CO_2$ desorption, then equilibrated at 300°C.

The variation of oxygen excess in Bi-2201 with temperature and $Po_2$ is presented in figure 2. Changes in composition were reproducible within 0.002 oxygen / formula unit.

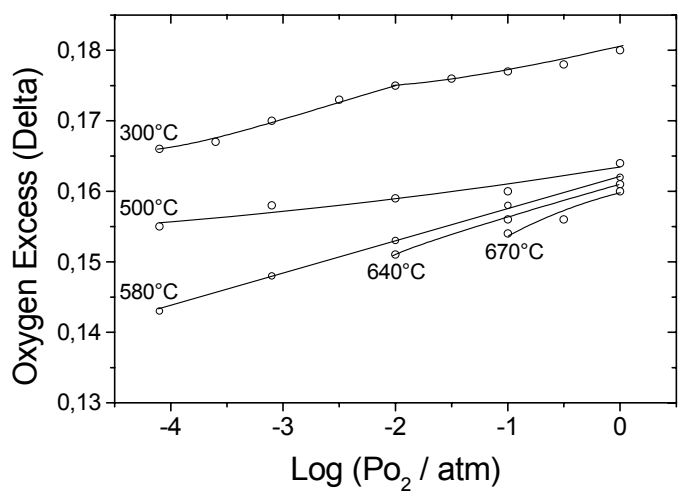

FIGURE 2
Isothermal variation of oxygen excess in Bi-2201, plotted at various temperatures.

It appears that oxygen excess in $Bi_2Sr_2CuO_{6+\Delta}$ varies within a small range, between 0.14 and 0.18 extra-oxygen per formula unit. The use of higher temperatures and/or lowest

Po$_2$ must be considered carefully because of phase decomposition[3]. Like in Bi-2212[6], the stoichiometric phase ($\Delta = 0$) is probably out of access, at least by the classical post-synthesis annealing method under atmospheric pressure. Another common feature with the two-layer analogue is the enhancement of oxygen excess dependence on Po$_2$ when temperature increases[4,6-8], associated with a modification of the isotherms shape. Inverted curvatures are observed whether T≤500°C or T>600°C, and a linear behaviour in the intermediate range (~580°C). This is likely to be related to different defect structures depending on the temperature used to adjust oxygen content, and is expected to affect significantly the oxide physical properties.

## 4. CONCLUSION

Controlling with accuracy the doping level of Bi-2201 through the oxygen excess implies the ability to prepare samples at regular intervals within the non-stoichiometric domain. In this view, systematic thermogravimetric measurements under well defined conditions of temperature and atmosphere composition were achieved. Oxygen excess in Bi$_2$Sr$_2$CuO$_{6+\Delta}$ can be adjusted in a reversible way between 0.14 and 0.18 by proper annealing. Our results also show that, as it was seen with Bi-2212, a specific regime of oxygen intercalation takes place in the high-temperature part of the diagram.